\documentclass[prl,twocolumn,groupedaddress,amsmath,floatfix]{revtex4}
\usepackage{graphicx}
\usepackage{bm}

\def\Kf{{k_{\rm F}}}

\newcommand{\coupling}{\gamma}

\newcommand{\vf}{v_{\rm F}}

\newcommand{\phdag}{{\phantom{\dagger}}}

\newcommand{\kf}{k_{\rm F}}

\newcommand{\txi}{\varepsilon}
\newcommand{\te}{E}
\newcommand{\dN}{{\cal{N}}}

\newcommand{\ef}{\epsilon_{\rm F}}

\newcommand{\bk}{{\bf k}}
\newcommand{\bq}{{\bf Q}}

\newcommand{\qh}{\hat{Q}}
\newcommand{\muh}{\hat{\mu}}

\newcommand{\Deltah}{\hat{\Delta}_Q}
\newcommand{\deltah}{\hat{\delta}}

\newcommand{\be}{\begin{equation}}
\newcommand{\ee}{\end{equation}}
\newcommand{\bea}{\begin{eqnarray}}
\newcommand{\eea}{\end{eqnarray}}
\newcommand{\bse}{\begin{subequations}}
\newcommand{\ese}{\end{subequations}}

\input{epsf}

\begin{document}
\title{BEC-BCS crossover in ``magnetized'' Feshbach-resonantly paired
  superfluids}
  
\author{Daniel E.~Sheehy and Leo Radzihovsky}
\affiliation{
Department of Physics, 
University of Colorado, 
Boulder, CO, 80309}
\date{August 18, 2005}
\begin{abstract}
  We map out the detuning-magnetization phase diagram for a
  ``magnetized'' (unequal number of atoms in two pairing hyperfine
  states) gas of fermionic atoms interacting via an s-wave Feshbach
  resonance (FR).
  The phase diagram
  is dominated by coexistence of a magnetized normal gas and a singlet
  paired superfluid with the latter exhibiting a BCS-Bose Einstein condensate crossover
  with reduced FR detuning. On the BCS side of strongly overlapping
  Cooper pairs, a sliver of finite-momentum paired
  Fulde-Ferrell-Larkin-Ovchinnikov magnetized phase intervenes between
  the phase separated and normal states. In contrast, for large
  negative detuning a uniform, polarized superfluid, that is a
  coherent mixture of singlet Bose-Einstein-condensed molecules and fully
  magnetized single-species Fermi-sea, is a stable ground state.
\end{abstract}
\maketitle
Recent experimental realizations of paired superfluidity in trapped
fermionic atoms interacting via a Feshbach resonance
(FR)~\cite{Regal,Zweirlien} have opened a new chapter of many-body atomic
physics.  Almost exclusively, the focus has been on {\em equal}
mixtures of two hyperfine states exhibiting pseudo-spin singlet
superfluidity
that can be tuned from the momentum-pairing BCS regime of strongly
overlapping Cooper pairs (for large positive detuning) to the
coordinate-space pairing Bose-Einstein condensate (BEC) 
regime of dilute molecules
(for negative detuning)~\cite{theory}.

In contrast, s-wave pairing for {\em unequal} numbers of atoms in the
two pairing hyperfine states 
has received virtually no experimental attention and 
only some recent theoretical activity\cite{Liu,Bedaque,Mizushima,Carlson,Pao,Son}.
Associating the two pairing hyperfine states with  up ($\uparrow$)
and down ($\downarrow$)  pseudo-spin $\sigma$,
the density difference $\delta n=n_\uparrow - n_\downarrow$ 
is isomorphic to ``magnetization'' $m\equiv\delta n$
and the corresponding chemical potential difference
$\delta\mu=\mu_\uparrow-\mu_\downarrow$ to a purely Zeeman field
$h\equiv\delta\mu/2$.

This subject dates back to the work of Fulde and Ferrell
(FF)~\cite{ff} and Larkin and Ovchinnikov (LO)~\cite{lo} who proposed
that, in the presence of a Zeeman field, an s-wave BCS superconductor
is unstable to magnetized pairing at a finite momentum $Q\approx
k_{{\rm F}\uparrow} -k_{{\rm F}\downarrow}$ 
with $ k_{{\rm F}\sigma}$ the Fermi wavevector of fermion $\sigma$.
This FFLO state, which remains elusive in condensed matter systems
where it is obscured by orbital and disorder effects,
spontaneously breaks rotational and translational symmetry and emerges
as a compromise between competing singlet pairing and Pauli
paramagnetism.

\begin{figure}
\epsfxsize=9.5cm
\centerline{\epsfbox{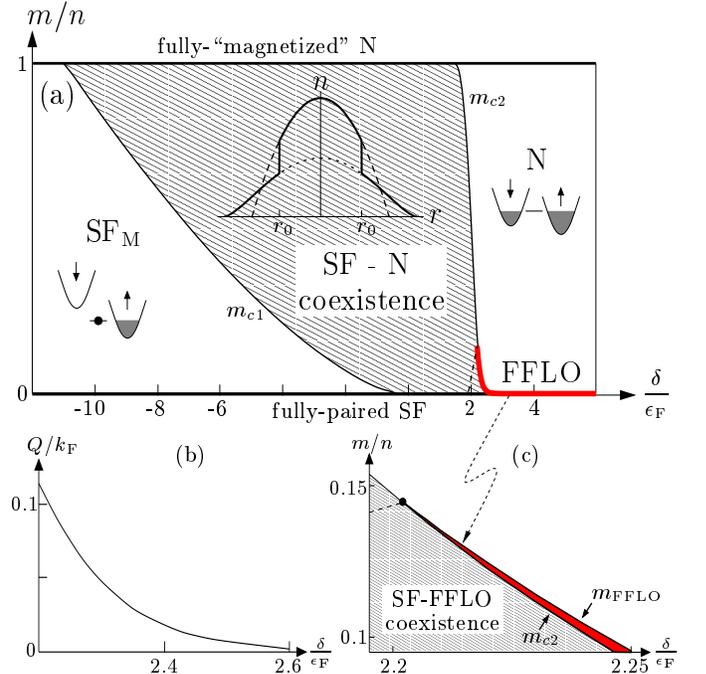}}
\vskip-.45cm
\caption{(Color Online) Detuning, $\delta$ -- population difference, $m/n =
  (n_\uparrow - n_\downarrow)/(n_\uparrow + n_\downarrow)$ phase
  diagram (for coupling $\coupling = 0.1$) in (a) displaying
  ``normal'' (N), magnetized superfluid (SF$_M$), FFLO (thick red line) and SF-N
  coexistence states, (b) showing the FFLO wavevector $Q(\delta)$ 
along the FFLO-N phase boundary,
  and (c) zoom-in on the FFLO state, stable only for $\delta>\delta_*\simeq 2.2 \ef$.
To the right of the dashed lines in (a) and (c), the SF-N coexistence undergoes 
a transition to SF-FFLO coexistence.
}
\label{fig:globalphase}
\vskip-.70cm
\end{figure}

Thus atomic fermion gases (where the above deleterious effects are
absent), tuned near an s-wave FR, are promising ideal systems for
a realization of the FFLO and related finite-magnetization paired
states, that can be studied throughout the full BCS-BEC crossover.

In this Letter, we map out the
detuning-magnetization phase diagram (Fig.\ref{fig:globalphase}) of 
such paired superfluids. We
find that for  positive detuning $\delta$ and arbitrarily small $m$, 
the system phase-separates into a
magnetized normal gas (N) and a singlet-paired BCS superfluid that
exhibits a BCS-BEC crossover with reduced $\delta$.  The FFLO state
intervenes in a sliver on the boundary between this coexistence region
and the N state.  For large negative detuning a uniform magnetized
superfluid (SF$_M$), that is a coherent mixture of singlet
Bose-condensed molecules and fully magnetized single-species Fermi-sea
is a stable ground state. Our predictions of these
states and transitions between them are testable
via thermodynamics (qualitatively modified by gapless atomic
excitations inside the SF$_M$ and FFLO states), sound propagation
(with zeroth sound velocity vanishing at the SF$_M$-N transition), and
 time of flight imaging (displaying density discontinuity and
striking Bragg peaks associated with the finite momentum pairing in
the FFLO state).

We now sketch the analysis that led to these results. A
gas of fermionic atoms,
$\hat{a}_{k\sigma}$, resonantly interacting through a diatomic
(closed-channel) molecule, $\hat{b}_q$, is described by a two-channel
Hamiltonian~\cite{theory}
\bea
\label{eq:hamiltonian}
&&H = \sum_{k,\sigma} (\epsilon_k - \mu_\sigma) \hat{a}_{k\sigma}^{\dagger}
\hat{a}_{k\sigma}^{\phdag} + \sum_q \big(\frac{\epsilon_q}{2} + \delta_0 -
2\mu)\hat{b}_q^\dagger \hat{b}_q^\phdag
\nonumber\\
&& \qquad + {g}\sum_{k,q}\Big(\hat{b}^\dagger_q
\hat{a}_{k+\frac{q}{2}\downarrow}^{\phdag}
\hat{a}_{-k+\frac{q}{2}\uparrow}^{\phdag}
+ h.c.
\Big), \eea
where $\epsilon_k \equiv k^2/2m_a$, $\mu_{\uparrow,\downarrow} = \mu
\pm h$ are the chemical potentials to impose atom number in hyperfine
states ${\uparrow,\downarrow}$, or the total atom density
$n=n_{\uparrow}+n_{\downarrow} + 2 \sum_q \langle
\hat{b}_q^\dagger \hat{b}_q\rangle$ (imposed by $\mu$, with
$n_\sigma=\sum_{q,\sigma} \langle \hat{a}_{k\sigma}^\dagger
\hat{a}_{k\sigma}\rangle$) and density difference (magnetization)
$m=n_\uparrow-n_\downarrow$ (imposed by $h$).  Here, $\delta_0$ is the bare
FR detuning, $g$ is the FR coupling determining the resonance width and
the system volume is unity.

For a narrow FR (small $g$), $H$ can be accurately analyzed by treating
$\hat{b}_q$ as a single-momentum\cite{ff,commentFF,Combescot,lo} c-number mode $\langle
\hat{b}_q\rangle \equiv b_Q\delta_{q,Q}$ with corrections small~\cite{smallnote} in
powers of $\coupling\equiv\gamma(\ef)= g^2 \dN(\ef)/\ef$, the ratio of
the FR width to Fermi energy, with $\dN(\ef) = 
m_a^{3/2}\sqrt{\ef}/\sqrt{2}\pi^2\equiv c\sqrt{\ef}$ the density of
states at the Fermi energy $\ef = \kf^2/2m$ set by the total atom
density $n = \frac{4}{3}c\ef^{3/2}$.  To lowest order in $\coupling$,
standard Bogoliubov analysis~\cite{longpaper} gives the ground-state
energy ($\hbar = 1$):
\bea &&E_G = \langle H\rangle = \big(\frac{\epsilon_Q}{2} +
\delta_0-2\mu \big)\frac{\Delta_{Q}^2}{g^2} - \sum_\bk (\te_k-\txi_k )
\nonumber \\
&& \qquad +\sum_\bk \big[
E_{\bk\uparrow} \Theta(-E_{\bk\uparrow})+
E_{\bk\downarrow} \Theta(-E_{\bk\downarrow})
\big],
\;\;\;\;\;
\label{eq:havg}
\eea
where $E_{\bk\sigma}=E_k\mp (h +\bk\cdot\bq/2m_a)$ is the excitation
spectrum for a hyperfine state $\sigma$, with ``gap'' $\Delta_Q \equiv
g b_Q$ and $E_k \equiv (\varepsilon_k^2 +\Delta_Q^2)^{1/2}$,
$\varepsilon_k \equiv \frac{k^2}{2m_a} - \mu + Q^2/8m_a$, and
$\Theta(x)$ the Heaviside step function. The corresponding ground
state is of the BCS form, but with pairing limited to momenta $\bk$
satisfying $E_{\bk\sigma} > 0$. 

The phase diagram is determined by minimizing $E_G$ over $Q$ and
$\Delta_Q$ at fixed average total density $n$, population difference
$m$, and physical detuning $\delta=\delta_0- g^2 \sum_k 1/2\epsilon_k$
(determined by the 2-body scattering amplitude). The competing ground
states are: (i) a normal Fermi gas (N) with $\Delta_Q=0$, (ii) a
non-\lq\lq magnetic\rq\rq\, fully paired BCS-BEC superfluid (SF) with
$\Delta_Q\neq 0$, $\bq=0$, and $m=0$, (iii) a magnetized partially
paired superfluid (SF$_M$) with $\Delta_Q\neq 0$, $\bq=0$, and $m\neq
0$, and (iv) a magnetized, finite-momentum paired superfluid
(FFLO) with $\Delta_Q\neq 0$, $\bq\neq 0$, and $m\neq 0$.
Anticipating the existence of first-order transitions, across which
 $m, n$ are discontinuous, in order to guarantee a solution
everywhere it is essential to also include phase-separated states
where two of above pure states coexist as a mixture in fractions $1-x$
and $x$ to be determined.

The computation of the ground-state energy is simplified by noting
that $E_G(h)=E_G(0)-\int_0^h m(h')d h'$, where $E_G(0)$ is the
well-studied fully-paired $h=0$ energy and $m(h)=-\partial
E_G/\partial h$
is the atom species imbalance number .
We compute $E_G$ by first neglecting the FFLO state (i.e., $Q=0$),
which, as we shall show, is only stable for a narrow window of
parameters (see Fig.~\ref{fig:globalphase}).
Then,
\bea
\label{three}
m(h)&=&\frac{2}{3}c
\Theta(h-\Delta)\big[(\mu+\sqrt{h^2-\Delta^2})^{3/2}\nonumber\\
&&-(\mu-\sqrt{h^2-\Delta^2})^{3/2}\Theta(\mu-\sqrt{h^2-\Delta^2})\big].
\;\;\;\;\;\;\;\;\eea
For {\it positive\/} detuning $\delta\gg\ef\gamma^{1/2}$, appropriate in
the BCS and throughout most of the crossover regimes, $\Delta \ll \mu$
and the density of states inside $E_G(0)$ can be well approximated by
a constant $\dN(\mu)$, giving:
\bea
&&\hspace{-0.5cm}
E^{+}_{G}\approx{1\over g^2}(\delta-2\mu)\Delta^2 +
\dN(\mu)\Big[-{1\over2}\Delta^2+\Delta^2\log{\Delta\over
  8e^{-2}\mu}\Big]\nonumber\\
&&\hspace{-0.3cm}
-8 \dN(\mu)\mu^2/15-\int_0^h m(h')d h'.
\label{Ebcs}
\eea
For small $h\ll \mu$ 
the species
imbalance contribution to $E_G$ is well approximated by $\int_0^h
m(h')d h'\approx \dN(\mu)\Theta(h-\Delta)\big[h\sqrt{h^2-\Delta^2}
-\Delta^2\cosh^{-1}(h/\Delta)\big]$. For $0 < h < \Delta_{BCS}/2$,
$E_G^{+}$ exhibits a single minimum at a standard ($h=0$) BCS value
$\Delta_{BCS}=8 e^{-2}\mu e^{-\coupling^{-1}(\delta-
  2\mu)/(\ef\mu)^{1/2}}$ and a maximum at $\Delta=0$. For a higher
Zeeman field $\Delta_{BCS}/2 < h < \Delta_{BCS}/\sqrt{2}$, the normal
state at $\Delta=0$ becomes a local minimum separated from the
$h$-independent global minimum at $\Delta_{BCS}$ by a maximum at
$\Delta_{Sarma}=\Delta_{BCS}\sqrt{2h/\Delta_{BCS}-1}$ \cite{Sarma}.
For $h > \Delta_{BCS}/\sqrt{2}$ the minimum at $\Delta=0$ lowers below
that of the BCS state. For a fixed $\mu$, this predicts a first-order
SF-N transition at $h_c(\mu,\delta)$ ,
with asymptotic
form in the narrow FR limit given by
\be 
h_c(\mu,\delta)\approx a_1\mu e^{-a_2\gamma^{-1}(\delta-
2\mu)/\sqrt{\ef\mu}},
\label{hc}
\ee
where $a_{1,2}=8e^{-2}/\sqrt{2}, 1\;\;(120^{2/5}e^{-8/5},4/5)$, for
$\mu\ll\delta/2$\;\;($\mu\gg\delta/2$). The transition is accompanied
by a jump in atom density from $n^{(S)}(\mu,\delta)\approx
{4\over3}\dN(\mu)\mu + 2 g^{-2}\Delta_{BCS}^2$ down to
$n^{(N)}(\mu,h_c)={2\over3}c\big[(\mu+h_c)^{3/2}+(\mu-h_c)^{3/2}\Theta(\mu-h_c)\big]
\approx n^{(S)}(\mu,\delta)-{4\over g^2}h_c^2(1-\gamma(\mu)/8)$, a
jump in species imbalance from $0$ to $m\approx
2\dN(\mu)h_c$, as well as other standard thermodynamic singularities.

In a more experimentally relevant ensemble of fixed total atom number
$n=-\partial E_G/\partial\mu$,
 for $h_{c1}\equiv h_c(\mu^{(S)}(n,\delta),\delta)< h < h_{c2}\equiv
h_c(\mu^{(N)}(n,h),\delta)$ neither SF nor N states can satisfy the
atom number constraint while remaining a ground state;
$\mu^{(S,N)}$ are SF and N chemical potentials at density
$n$, Zeeman field $h$ and detuning $\delta$, obtained by solving
$n=n^{(S,N)}(\mu^{(S,N)})$ above. For a narrow FR, $\gamma\ll 1$, we find
\bea
\label{six}
\hspace{-0.7cm}
h_{c1}(\delta,n)&\approx&
\begin{cases}
\frac{1}{\sqrt{2}}\Delta_F(\delta)
e^{-\big({\delta\Delta_F(\delta)\over2\sqrt{2}\gamma\epsilon_F^2}\big)^2},
& \text{for $\delta\gg 2\ef$,}\cr
\frac{1}{2}g\big[n-\frac{4}{3}c(\delta/2)^{3/2}\big]^{1/2},
& \text{for $\delta\ll 2\ef$,}\cr
\end{cases}\\
\hspace{-0.7cm}
h_{c2}(\delta,n)&\approx&
\begin{cases}
\frac{1}{\sqrt{2}}\Delta_F(\delta)
e^{-{\delta\over 16\gamma\epsilon_F}\big({\Delta_F(\delta)\over\ef}\big)^2},
& \text{for $\delta\gg 2\ef$,}\cr
h^{(N)}(\delta/2,n),
& \text{for $\delta\ll 2\ef$},\cr
\end{cases}
\label{hc1hc2}
\eea
where $\Delta_F\equiv\Delta_{BCS}(\delta,\ef)$ and
$h^{(N)}(\delta/2,n)$ is the solution of
$n=n^{(N)}(\delta/2,h^{(N)})$.  Hence for $h_{c1} < h < h_{c2}$ the
gas phase separates~\cite{Bedaque} into SF and N rich regions in
$1-x(h,\delta)$ and $x(h,\delta)$ proportions, determined by the atom
number constraint $x n^{(N)} + (1-x) n^{(S)} = n$. In above
$n^{(S)}(\mu_c,\delta) > n > n^{(N)}(\mu_c,\delta)$ are the SF and N
state densities computed along the critical chemical potential
$\mu_c(h,\delta)$ determined by Eq.~(\ref{hc}) with limiting values
$\mu_c(h_{c1,2},\delta)=\mu^{(S,N)}(n,\delta)$.  The fraction of the N
state admixture is then given by
$x(h,\delta)=\left(n^{(S)}(\mu_c,\delta)-n\right)
\left[n^{(S)}(\mu_c,\delta)-n^{(N)}(\mu_c,\delta)\right]^{-1}$
ranging between $0$ and $1$ for $h_{c1} < h < h_{c2}$ spanning the
coexistence region. 

A single-valued relation between the magnetization (species imbalance)
$m(h,\delta)=\frac{2}{3}c\big[(\mu_c(h,\delta)+h)^{3/2}-
(\mu_c(h,\delta)-h)^{3/2}\Theta(\mu_c-h)\big]$ and Zeeman field $h$ in
the normal paramagnetic state allows us to reexpress above predictions
in terms the species imbalance number $M = x m$, that is, the
quantity (rather than $h$) that we anticipate to be kept fixed in
atomic gas experiments.  As illustrated in Fig.\ref{fig:globalphase}
in a phase diagram expressed in terms of $m\equiv\delta n$ the
fully-paired SF state is confined to the detuning axis ($m=0$) and the
boundary between the coexistence region and the N state is given by
$m_{c2}(n,\delta)\equiv m(h_{c2},\delta)$.

We now turn to the {\it negative\/} detuning (BEC) regime.  Although $E_G$,
Eq.~(\ref{eq:havg}) and the phase diagram that follows from it can be
accurately computed numerically (Fig.\ref{fig:globalphase}),
considerable insight can be gained by analytical analysis. This is
particularly simple in the $\gamma\rightarrow0$ limit in which
$E_G^{-}\approx (\delta-2\mu)|b|^2 -
(4c/15)(h+\mu)^{5/2}\Theta(h+\mu)$,
$m(h,\mu)=(2c/3)(h+\mu)^{3/2}\Theta(h+\mu)$, and $n = 2|b|^2 + m$.

For $h=0$ and $\delta < 0$, this shows that the BCS superfluid
ground-state smoothly crosses over to a BEC of closed-channel
molecules, with a finite atom excitation gap 
$\sqrt{\mu^2+g^2|b|^2}\approx |\mu|$ enforcing atom vacuum and the
condensate density $|b|^2\approx n/2+{\cal{O}(\gamma)}$. The gap
equation $\partial E_G/\partial b=2(\delta-2\mu)b\approx 0$ then
determines $\mu\approx\delta/2$, as in the crossover region,
$\ef\gamma^{1/2}<\delta<2\ef$, above. From the excitation spectrum
$E_{k\sigma}$ it is clear that this ground state remains stable for
$0<h<h_{m}(\delta)$, with $h_{m}(\delta)=\sqrt{g^2 n/2 +
  (\delta/2)^2}\approx -\delta/2$ determined by
$E_{0,\uparrow}(\delta,h_{m})=0$. However, for $h > h_{m}$, finite
species imbalance $m\approx(2c/3)(h-|\delta|/2)^{3/2}$
develops, depleting the SF condensate $|b|^2=n/2 -
(c/3)(h-|\delta|/2)^{3/2}$. The resulting magnetized SF$_M$ is 
stable for $h_{m} < h < h_{c2}$, with $h_{c2}(\delta)=(3n/2
c)^{2/3}+|\delta|/2=2^{2/3}\ef-\delta/2$ determined by
$m(h_{c2},\delta/2)=n$, giving a smooth extension into the BEC regime
of Eq.~(\ref{hc1hc2}), computed inside the BCS and crossover regimes.

\begin{figure}
\epsfxsize=9.0cm
\centerline{\epsfbox{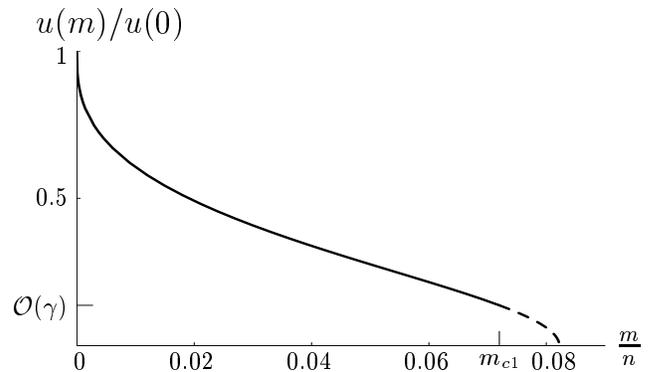}}
\vskip-.7cm
\caption{
  Bogoliubov sound speed $u$ in the BEC regime as a function of
  species imbalance $m$ for $\delta = -2\ef$, vanishing at
  boundary of the SF$_M$ with the $SF-N$ coexistence region.}
\label{velocityplot}
\vskip-.5cm
\end{figure}

For a narrow FR $E^-_G(\Delta,\mu,\delta,h)$ can be accurately computed
analytically giving
\bea
&&\hspace{-0.5cm}
E^{-}_{G}\approx{1\over g^2}(\delta-2\mu)\Delta^2 +
c|\mu|^{\frac{5}{2}}\frac{\pi}{2}\Big[\left(\frac{\Delta}{|\mu|}\right)^2
+\frac{1}{2^5}\left(\frac{\Delta}{|\mu|}\right)^4\nonumber\\
&&-\frac{5}{2^{10}}\left(\frac{\Delta}{|\mu|}\right)^6+
\frac{105}{2^{16}}\left(\frac{\Delta}{|\mu|}\right)^8\Big]
-\int_0^h m(h')d h'.
\label{Ebec}
\eea
Its minimization together with the atom number constraint fixes
$\mu\approx\delta/2 +  {\cal O}(\gamma)$
 and
leads to the phase diagram in Fig.~\ref{fig:globalphase}.  We
find\cite{longpaper} that above expressions for $h_m(\delta)$ and
$h_{c2}(\delta)$ receive only small ${\cal O}(\gamma)$ correction for
$\delta < -\gamma^{1/2}\ef$. Hence in contrast to the BCS side, where
the system undergoes  phase separation for an arbitrary small 
$m \neq 0$, on the BEC side the transition at
$h_m(\delta)$ is into a uniform magnetized superfluid (SF$_M$) that
persists over a finite range of $m$
  and is a coherent superposition of a singlet molecular condensate 
and fully spin-polarized
Fermi gas.
The sequence
SF$\rightarrow$ SF$_M\rightarrow$ N of continuous
transitions remains unchanged for $\delta <
\delta_c\approx-10.6\ef$.  However, for a finite $\gamma$ and
$\delta_c\approx-10.6\ef < \delta < -\gamma^{1/2}\ef$ a secondary
local (N state) minimum develops at $\Delta=0$ leading to a 1st-order
SF$_M\rightarrow$N transition at $h_{c1}(\delta)\approx -0.65\delta <
h_{c2}(\delta)$, preempting a continuous one at $h_{c2}(\delta)$.  For
a fixed atom density $n$ and $h_{c1}(\delta) < h < h_{c2}(\delta)$ the
gas phase-separates into coexisting SF$_M$ and $N$ states.

This $h_{c1}(\delta)$ boundary (equivalent to 
$m_{c1}(\delta)\equiv m(h_{c1}(\delta),\delta/2)\approx 0.029 n
|\delta/\ef|^{3/2}$) is accurately (to ${\cal O}(\gamma^2)$) located
by the vanishing of the coefficient of $\Delta^4$ in $E_G^-(\Delta)$,
proportional to the effective molecular scattering length
$a_m(h,\delta)=
\frac{\sqrt{2}\gamma^2 \hbar\ef \pi^2}{64 \sqrt{m_a}|\mu|^{3/2}}F(h/|\mu|)$.
  We
then predict\cite{longpaper} that the Bogoliubov sound velocity
$u(\delta,m)$ vanishes (to ${\cal O}(\gamma)$, followed by a small jump to
$0$) at the 1st-order SF$_M$-N transition and exhibits a $-m^{1/3}$
cusp singularity at the SF-SF$_M$ boundary. The full
expression is illustrated in Fig.~\ref{velocityplot} and given by 
(with $\deltah \equiv \delta/\ef$)
\be 
u \approx u_0 \sqrt{1-m/n}\sqrt{F\Big(1+\frac{2^{5/3}}{|\deltah|} 
\Big(\frac{m}{n}\Big)^{2/3}\Big)}, 
\ee
with $F(x)\equiv 1- \frac{2}{\pi x^2}\big[\sqrt{x-1}(x+2)
+x^2\tan^{-1}\sqrt{x-1}\big]$ and $u_0 = \frac{2^{3/4}\gamma}{8\sqrt{3}} 
\frac{\vf\sqrt{\pi}}{|\deltah|^{3/4}}$ the sound
velocity at $m=0$.

We now turn to the FFLO state. Because ${\bf Q}\neq 0$ pairing is
driven by the mismatch of the up $\uparrow$ and $\downarrow$ Fermi
surfaces, with $Q\approx \Kf_{\uparrow}-\Kf_{\downarrow}$, it is clear
that the FFLO state can only be stable at large positive detuning.
Computing $E_G^+$ for ${\bf Q}\neq 0$ to leading order in
${\Delta}_Q\ll\mu$ (using dimensionless quantities 
 $\Deltah \equiv \Delta_Q/\ef$, $\muh \equiv \mu/\ef$
$\hat{h} \equiv h/\ef$, $\qh = Q\sqrt{\muh}/\kf$, and
$\varepsilon_G\equiv E_{G}/c\ef^{5/2}$), we find:
\bea
\label{eq:eg}
&&\hspace{-0.4cm}
\varepsilon_G \approx  -\frac{8}{15}\muh^{5/2}
+\frac{ \qh^2 \Deltah^2}{2\coupling\muh}  +\sqrt{\muh}\Big[-\Deltah^2 - \hat{h}^2
 \\
&&\hspace{-0.4cm}+
\frac{\Deltah^2}{2} \ln \frac{4(\qh+\hat{h})(\qh-\hat{h})}{\hat{\Delta}_{BCS}^2} 
+ \frac{h \Deltah^2}{2\qh} \ln \frac{\qh+\hat{h}}{\qh-\hat{h}} 
+ \frac{\Deltah^4/8}{\qh^2-\hat{h}^2}\nonumber
\Big].
\eea
At fixed $\mu$, for a given $\deltah$ and $\hat{h}$, the ground state is
determined by minimizing $\varepsilon_G(\Deltah,\qh)$ over $\Deltah$
(the gap equation) and $\qh$ (equivalent to vanishing of the ground
state momentum).
For $\delta\gtrsim 2\ef$ we find a 1st-order SF-FFLO (preempting SF-N)
transition approximately at $h_c(\delta,\mu)$, Eq.~(\ref{hc}). At fixed
atom number, for $h>h_c$ the gas phase separates into coexisting SF
and FFLO states, approximately bounded above by $h_{c2}(\delta)$,
computed for $Q=0$ above. At slightly higher field,
$h_{\rm FFLO}(\delta)$,
 we find that the FFLO state undergoes a
continuous transition (that on general grounds we expect to be driven
1st-order by fluctuations) into the N state. Numerical solution of the
gap, number and momentum equations yields $h_{FFLO}(\delta)$ (and thus
  $m_{\rm FFLO}$ via Eq.~(\ref{three}), plotted in Fig.1)
that interpolates
between $0.754 \Delta_F(\delta)$ for large $\delta$ (in agreement with 
FF~\cite{ff}) and $h_{c1}(\delta,n)$ for $\delta \to \delta_*$, 
with the crossing point $\delta_* \approx 2\ef$.
This microscopically calculated value of $\delta^*>0$ contrasts with
the conclusion of Ref.~\onlinecite{Son}, the latter based on a purely
{\em qualitative} discussion, that has little {\it quantitative\/}
predictive power, e.g., in determining the precise location of
phases.

In free expansion experiments, the FFLO state, 
most easily observed with a trap having a typical size that is large
compared to $Q^{-1}$,
should exhibit a BEC
peak (observed by its projection onto the molecular
condensate\cite{Regal}) shifted by $\hbar{\bf Q}t/m_a$ ($t$
expansion time) corresponding to the finite momentum $\hbar{\bf Q}(\delta)$
(Fig.\ref{fig:globalphase}(b)) of its condensate, and a (spontaneous)
Bragg lattice of peaks in the more-likely case of  multiple-${\bf Q}$
pairing\cite{lo,commentFF,Combescot,Bowers}. The anisotropy of the FFLO pairing should
also be reflected in ``noise'' experiments\cite{Greiner05} sensitive
to {\it angle-dependence\/} of pairing correlations across the Fermi
surface. Our predictions of gapless atomic excitations in the SF$_M$
and FFLO states, as well as the vanishing of 
the molecular scattering length $a_m$ and of 
the 0th sound velocity
$u$ at the SF$_M$-N phase boundary should be observable through Bragg
spectroscopy and reflected in  thermodynamics (e.g., heat capacity
that is power-law in $T$).  We also expect standard thermodynamic
anomalies across phase transitions in Fig.\ref{fig:globalphase}(a),
and phase separation accompanied by density discontinuity and local
density variation with detuning and atom imbalance across the
coexistence region.  Finally, because a gas trapped in a smooth
potential $V(r)$ is well-characterized by a local chemical potential
$\mu(r)\equiv\mu - V(r)$ (the Thomas-Fermi approximation), our fixed
$\mu$ analysis is directly experimentally relevant. For negative
detuning and finite species imbalance we predict SF state in the
cloud's core of radius $r_0(\delta)$, with density discontinuity to
the outer-shell N state, with $r_0(\delta)$ determined by
$\mu(r_0)=\mu_c(\delta,m)$ (see center inset of
Fig.\ref{fig:globalphase}(a)).  We expect that this shell structure
should be readily observable, particularly if different hyperfine
states and closed and open channels can be imaged independently.
Details of these experimental predictions will be presented in a
forthcoming publication\cite{longpaper}.

\noindent
We  acknowledge discussions with V. Gurarie, D. Jin, E. Mueller
and  C. Regal as well as support from NSF DMR-0321848 and the Packard
Foundation.


\begin{thebibliography}{10}
\bibitem{Regal}
C.A. Regal, M. Greiner, and D.S. Jin, 
%
%
Phys. Rev. Lett. {\bf 92}, 040403 (2004).
%
%
\bibitem{Zweirlien}
M.W. Zwierlein {\it et al.\/}, 
%
%
Phys. Rev. Lett. {\bf 92}, 120403 (2004).
%
%
\bibitem{theory}
E. Timmermans {\it et al.\/}, 
Phys. Lett. A {\bf 285}, 228 (2001);
%
M. Holland {\it et al.\/}, Phys. Rev. Lett. {\bf 87}, 120406 (2001).
\bibitem{Liu}
W. V. Liu and F. Wilczek,
Phys. Rev. Lett. {\bf 90}, 047002 (2003).
\bibitem{Bedaque}
P.F. Bedaque,
 H. Caldas, and G. Rupak,
Phys. Rev. Lett. {\bf 91}, 247002 (2003).
%
\bibitem{Mizushima}
 T. Mizushima,
K. Machida, and M. Ichioka,
Phys. Rev. Lett. {\bf 94}, 060404 (2005).
\bibitem{Carlson}
J. Carlson and S. Reddy,
Phys. Rev. Lett. {\bf 95\/}, 060401 (2005).
\bibitem{Pao}  C.-H. Pao  {\it et al.\/}, cond-mat/0506437.
\bibitem{Son}
D. T. Son and M.A. Stephanov, cond-mat/0507586.
\bibitem{ff}
P. Fulde and R. A. Ferrell,
Phys. Rev. {\bf 135}, A550 (1964).
\bibitem{lo}
A.I. Larkin and Yu.N. Ovchinnikov,
Zh. Eksp. Teor. Fiz {\bf 47}, 1136 (1964) 
[Sov. Phys. JETP {\bf 20}, 762 (1965)].
\bibitem{commentFF} We avoid the complexity of
{\em multiple} {\bf Q} FFLO states,  known to be very close in energy\cite{lo,Combescot},
 therefore not
  affecting the global phase diagram.
\bibitem{Combescot}   
R. Combescot and C. Mora,
Europhys. Lett. {\bf 68}, 79 (2004).
%
\bibitem{smallnote}
Our results are {\it quantitatively\/} valid for narrow resonance (small $\gamma$);
we have also qualitatively reproduced them within a wide-resonance single-channel model~\cite{longpaper}.
\bibitem{longpaper}
D.E. Sheehy and L. Radzihovsky,  {\it in preparation\/}.
\bibitem{Sarma}
G. Sarma, J. Phys. Chem. Solids {\bf 24}, 1029 (1963).
%
\bibitem{Bowers}
 J. A. Bowers and K. Rajagopal,
Phys. Rev. D {\bf 66}, 065002 (2002).
\bibitem{Greiner05}
M. Greiner {\it et al.\/}, Phys. Rev. Lett. {\bf 94}, 110401 (2005). 
%
\end{thebibliography}
\end{document}